\documentclass[preprintnumbers,amsmath,amssymbm,prl]{revtex4}
\usepackage{epsfig}
\usepackage{graphicx}

\begin{document}
\title{Comment on ``The extremal black hole bomb"}
\author{Shahar Hod}
\affiliation{The Ruppin Academic Center, Emeq Hefer 40250, Israel}
\affiliation{ }
\affiliation{The Hadassah Institute, Jerusalem 91010, Israel}
\author{Oded Hod}
\affiliation{The Sackler Faculty of Exact Sciences, Tel Aviv University, Tel Aviv 69978, Israel}
\date{\today}

\begin{abstract}
\ \ \ Recently, we have provided an analytical treatment of the
phenomena known as the 'black-hole bomb' (arXiv:0910.0734). In
particular, we have determined analytically the unstable growing
resonances of a massive scalar field in the rotating Kerr black hole
spacetime. It was later claimed by J. G. Rosa (arXiv:0912.1780) that
the analytic procedure may fail for some values of the field's mass.
This claim was based on the concern that some of the Gamma functions
that are involved in the analysis may develop poles. In this comment
we show, by explicit calculations, that the Gamma functions, which are
used, are all well behaved near the peak of the black-hole
resonances. This fact supports the validity of our previous analytical
treatment. We further comment on the regime of validity of some of the
approximations used by Rosa.
\end{abstract}
\bigskip
\maketitle


A bosonic field impinging on a rotating black hole can be amplified
as it scatters off the hole, a phenomena known as superradiant
scattering. If in addition the field has a non-zero rest mass then
the mass term effectively works as a mirror, reflecting the
scattered wave back towards the black hole. In this physical system,
known as a black-hole bomb \cite{PressTeu2,Car}, the wave may bounce
back and forth between the black hole and some turning point
amplifying itself each time. Consequently, the massive field grows
exponentially over time and is unstable.

Former analytical estimates of the instability timescale associated
with the dynamics a of massive scalar field in the rotating Kerr
spacetime were restricted to the regimes $M\mu\gg 1$ \cite{Zour} and
$M\mu\ll 1$ \cite{Det,Furu}, where $M$ and $\mu$ are the masses of
the black hole and the field, respectively. In these two limits the
growth rate of the field (the imaginary part $\omega_I$ of the
mode's frequency) was found to be extremely weak. However,
subsequent numerical investigations \cite{Zour,Furu,Dolan} have
indicated that the instability is actually greatest in the regime
$M\mu=O(1)$, where the previous analytical approximations break
down. Thus, a new analytical study of the instability timescale for
the case $M\mu=O(1)$ is physically well motivated.

In a recent paper \cite{HodHod1} we have studied {\it analytically}
for the first time the phenomena of superradiant instability (the
black-hole bomb mechanism) in the physically interesting regime
$M\mu=O(1)$, the regime of greatest instability. We have shown that
the resonance condition for the bound states of the field is given
by
\begin{equation}\label{Eq19}
{1\over{\Gamma({1\over
2}+\beta-\kappa)}}=(8i)^{2\beta}\Big[{{\Gamma(-2\beta)}\over{\Gamma(2\beta)}}\Big]^2{{\Gamma({1\over
2}+\beta-ik)}\over{\Gamma({1\over 2}-\beta-ik)\Gamma({1\over
2}-\beta-\kappa)}}\Big[Mr_+\sqrt{\mu^2-\omega^2}(m\Omega-\omega)\Big]^{2\beta}\
,
\end{equation}
see Ref. \cite{HodHod1} for details. [We use here the same notations
as in Ref. \cite{HodHod1}]. Solving this resonance condition yielded
an instability growth rate of $\tau^{-1}\equiv\omega_I=1.7\times
10^{-3}M^{-1}$ for the fastest growing mode \cite{HodHod1}. This
instability is four orders of magnitude {\it stronger} than has been
previously estimated \cite{Dolan}.

\begin{figure}[h]
  \begin{center}
    \epsfxsize=10.0cm \epsffile{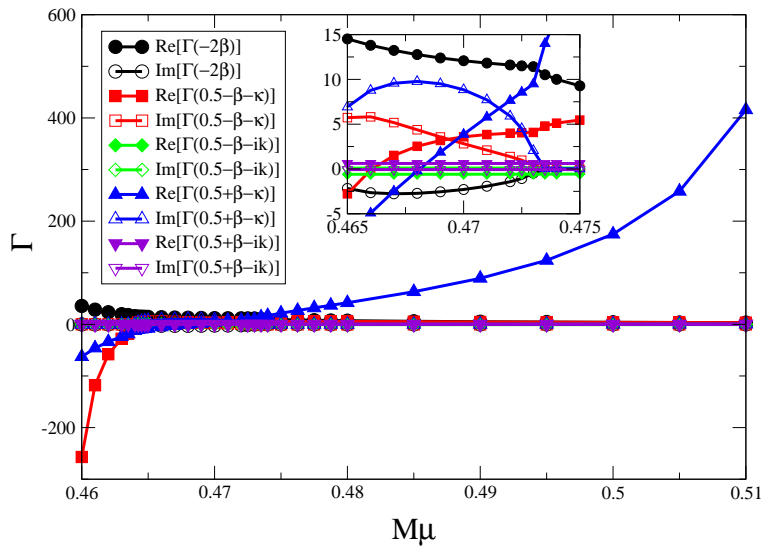}
  \end{center}
  \caption
      {The behavior of the various Gamma functions as a function of
  the field's mass $\mu$.  The results are for the $l=m=1$ mode, the
  mode with the greatest instability. Inset: zoom-in on the region
  where the largest resonance is obtained. The Gamma functions are all found
  to be well behaved near the spectrum's peak.}
      \label{Fig1}
\end{figure}

It was recently claimed \cite{Rosa} that the analytic treatment of
Ref. \cite{HodHod1} may fail for some values of the field's mass.
This claim was based on the concern that some of the Gamma functions
that are involved in the analysis may develop poles. In order to
remove this concern, we explicitly calculated all the Gamma
functions which are involved in the analysis [see Eq. (\ref{Eq19})].
We depict the results in figure \ref{Fig1}. It is clear that the
Gamma functions are all well behaved in the vicinity of the peak
black-hole resonances. Moreover, we note that at the edges of the
presented mass spectrum, where some of the Gamma functions are
largest, our analytical results do agree with the suggested
treatment of \cite{Rosa}, see Fig. 3a. of \cite{Rosa}. This clearly
supports the validity of our results near the peak of the spectrum
(at $M\mu\simeq 0.469$), where the Gamma functions are actually
smaller (that is, further away from their poles).

\input{epsf}

It is worth noting that Strafuss and Khanna \cite{Khanna} have used an
independent numerical technique to evaluate the instability growth
rate of a massive scalar field in the spacetime of a near extremal
rotating black hole. They have found an instability growth rate of
$\tau^{-1}\equiv\omega_I=2\times 10^{-5}M^{-1}$ for the case
$M\mu=0.25$. This is two orders of magnitude stronger than the
estimate of \cite{Dolan}, a fact which indicates that the instability
found in \cite{Dolan} is actually not the largest one possible. Since
the black-hole instability is expected to be more pronounced in the
regime $M\mu\simeq 0.5$ that we have analyzed in \cite{HodHod1} (as
compared to the $M\mu=0.25$ case that was studied in \cite{Khanna}),
one may consider the result of \cite{Khanna} as a lower bound on the
strength of the instability for near-extremal black holes. The
analytical treatment of \cite{HodHod1} reveals that this expectation
is indeed correct.

Finally, we would like to take this opportunity to remark on the
regime of validity of the analytical treatment used in \cite{Rosa}.
The validity of the analytical method, first developed in
\cite{HodHod1}, is restricted to the regime $\omega\simeq m\Omega
\simeq \mu$ with $\Omega\simeq 1/2M$ \cite{NoteAlm}. (Fortunately,
this is the regime of physical interest, where the instability is
greatest.) For the most unstable mode with $m=1$ this implies that
the analytical treatment is valid in the regime $M\mu\simeq {1\over
2}$. Thus, using the analytical approximation down to $M\mu\simeq
0.2$ as was done in \cite{Rosa} is actually physically not
justified.

In summary, we have shown that the Gamma functions involved in the
analysis of the black-hole bomb phenomena are all well behaved at
the peak resonances. This fact supports our previous findings
\cite{HodHod1} regrading the instability timescale associated with
the dynamics of a massive scalar field in the Kerr spacetime. Our
findings are further supported by the independent numerical analysis
performed in \cite{Khanna}.

\bigskip
\noindent
{\bf ACKNOWLEDGMENTS}
\bigskip

This research is supported by the Meltzer Science Foundation. We thank
Liran Shimshi, Clovis Hopman, Yael Oren, Adi Zalckvar, Ophir Ariel,
and Arbel M. Ongo for helpful discussions.



\begin{thebibliography}{99}
\bibitem{NoteAlm} The requirement $\omega\simeq\mu$ is needed in order to treat
$a^2(\omega^2-\mu^2)$ as a perturbation term in the equation that
determines the values of the seperation constants $A_{lm}$
\cite{HodHod1}. Under this condition one can then expand the
separation constants $A_{lm}$ in powers of $a^2(\mu^2-\omega^2)$ to
find $A=l(l+1)+\sum_{k=1}^{\infty}c_ka^{2k}(\mu^2-\omega^2)^k$. The
expansion coefficients $\{c_k\}$ are given in \cite{Abram}. [For
example, for $l=m=1$, the case of physical interest, we find
$c_1=1/5,\ c_2=-4/875,\ c_3=8/65625,...$. Note that in
\cite{HodHod1} we had a typo in the value of $c_1$, tough our
calculations in \cite{HodHod1} were performed with the correct value
of that coefficient].
\end{thebibliography}
\end{document}